
\input phyzzx

\hoffset=0.375in
\overfullrule=0pt
\def\br{{\bf r}}
\def\kms{{\rm km}\ {\rm s}^{-1}}
\def\kpc{\rm kpc}
\def\ms{{MACHOs}}
\def\m{{MACHO}}
\def\wc{{OGLE}}
\def\lmc{{\scriptscriptstyle\rm LMC}}
\def\smc{{\scriptscriptstyle\rm SMC}}
\def\bw{{\scriptscriptstyle\rm BW}}
\def\tlmc{{$\tau_{\scriptscriptstyle\rm LMC}$}}
\def\tsmc{{$\tau_{\scriptscriptstyle\rm SMC}$}}
\def\tsmctlmc{{$\tau_{\scriptscriptstyle\rm SMC}/\tau_{\scriptscriptstyle\rm
LMC}$}}
\def\eff{{\rm eff}}
\def\ev{{\rm events}}
\def\obs{{\rm obs}}
\def\spike{{\rm spike}}
\def\min{{\rm min}}
\def\max{{\rm max}}

\def\deg{^{\rm o}}

\def\kms{\hbox{ km s$^{-1}$}}
\def\kpc{\hbox{ kpc }}

\def\lsim{ \rlap{\lower .5ex \hbox{$\sim$} }{\raise .4ex \hbox{$<$} } }
\def\gsim{ \rlap{\lower .5ex \hbox{$\sim$} }{\raise .4ex \hbox{$>$} } }
\def\solar{ {\odot} }

\def\msolar{ {M_{\odot}} }

\def\cf{{cf.~\/}}
\def\eg{{e.g.,~\/}}
\def\ie{{i.e.,~\/}}
\def\etal{{et~al.~\/}}

\font\bigfont=cmr17
\centerline{\bigfont MACHOs in a Flattened Halo}
\bigskip
\centerline{\bf Penny D.\ Sackett and Andrew Gould}
\bigskip
\centerline{Institute for Advanced Study, Princeton, NJ 08540}
\medskip
\centerline{E-mail: psackett@guinness.ias.edu; gould@sns.ias.edu}
\bigskip
\singlespace
\centerline{\bf ABSTRACT}

If massive compact halo objects (\ms) are detected in
ongoing searches, then
\tsmctlmc,
the ratio of the optical depth toward the
Small and Large Magellanic Clouds, will be a robust indicator of
the flattening of the Galactic dark matter halo. For a spherical halo,
\tsmctlmc\ is about 1.45, independent of details of the
shape of the Galactic rotation curve, the assumed mass of the
Galactic disk and spheroid, and the truncation distance (if any)
of the dark halo.
For an E6 halo (axis ratio $c/a = 0.4$), the ratio of optical
depths is
\tsmctlmc\ $\sim 0.95$, again independent of assumptions
about Galactic parameters.  This ratio can be measured with
a precision as good as $\sim 10\%$ depending on the typical mass
of the MACHOs.
If the halo is highly flattened (\eg E6) and
closely truncated (\eg at twice the solar galactocentric
radius), then the optical depth toward the LMC can be reduced by a factor
of about two.
For these extreme parameters, and the assumption of a heavy Galactic
disk and spheroid, the upper limit of
the \m\ mass range
to which ongoing experiments are sensitive is reduced from
${\cal O}(10^6)\,M_\odot$ to ${\cal O}(10)\,M_\odot$.

\vskip 1.75in

Subject Headings:  gravitational lensing, dark matter, Magellanic Clouds
\endpage
\normalspace

\chapter{Introduction}

Three experiments are currently underway to detect Massive
Compact Halo Objects (\ms) in the Galactic halo using a method
originally suggested by Paczy\'nski (1986).  The \m\
Collaboration (Alcock \etal 1992) is now monitoring
$\sim 10^7$ stars in the Magellanic Clouds to search for the
characteristic microlensing light curve induced by the passing
of a \m\ across the line of sight to the source.  The
program stars are mostly in the Large Magellanic Cloud (LMC), with
some in the Small Magellanic Cloud (SMC).  During the
southern winter when the LMC is down, the \m\ Collaboration
plans to observe several million stars in the Galactic bulge.
The experiment is scheduled to last four years.
The Warsaw-Carnegie-Princeton Optical Gravitational Lens
Experiment (\wc) is observing several million stars in the
Galactic bulge over several years, and has completed one season of
observations (Udalski 1992).  A French collaboration is using
Schmidt plates and CCD frames to look for microlensing in LMC fields
(Aubourg \etal 1991).

Paczy\'nski (1991) and Griest \etal (1991) have suggested that
by observing the bulge one could measure
the low-mass end of the disk luminosity function and any baryonic dark
matter that may be present in the Galactic disk.
The bulge observations will also
be sensitive to \ms\ , although as noted by Griest \etal,
assuming standard halo parameters the optical depth
to microlensing by \ms\
in this direction is only $\sim 1/3$ the optical depth
toward the LMC and also only $\sim 1/3$ the optical
depth toward the bulge due to known disk stars.
However, as
Griest \etal pointed out, the optical depth toward the
bulge would be enhanced if the halo were flattened.

Here we consider the general effect of halo
flattening on the detectability of \ms\
by addressing two questions:
(1) If \ms\ are detected,
is there any specific signature that they are distributed
in a flattened rather than spherical halo?
(2) If \ms\ are not detected, could this negative result be attributed
to the \ms\ being in a flattened halo?

At first sight these questions appear difficult to
answer because
the core radius, asymptotic speed, and truncation radius of the
``standard'' spherical isothermal model for the dark halo
are not well constrained,
so that adding a fourth poorly-known parameter (halo flattening) would
seem to render the problem nearly insoluble.
We will show, however, that there are
signatures along certain lines of sight for which the optical depth of a
flattened versus spherical halo are largely independent of
these considerations.

The standard spherical halo is given by
$$\rho(\br) = {v_\infty^2\over 4\pi G}\,\biggl( {1 \over {a^2 + r^2}} \biggr)
\theta(R_T - r),\eqn\rhoofr$$
where $r$ is the Galactocentric radius,
$v_\infty$ is the asymptotic circular speed of the halo (assuming that
it extends to infinity); $a$ is the core radius
of the halo; and $R_T$ is the truncation radius.  This form can be
generalized to include a family of oblate halos of arbitrary axis ratio
(Sackett \& Sparke 1990, and \S\ 2).

As we discuss in \S\ 3, the circular speed and the core radius
are partially constrained by observations, while the truncation radius is
largely unknown.
We will therefore consider separately the effect of flattening on
truncated and untruncated
halos (or halos truncated only beyond the Magellanic Clouds).

Untruncated halos described by equation \rhoofr\ have only two
parameters, $a$ and $v_\infty$, both of which
are constrained by the observed rotation curve
and the assumed masses and mass distributions of the Galactic disk and
spheroid.  The Galactic rotation curve
measures
the total gravitational force
in the plane as a function of Galactocentric
distance; the assumed disk and spheroid models tell us what
fraction of this force is accounted for by the Galaxy's
observed components, and hence what fraction
is contributed by the dark halo.
The principal uncertainties are the slope of the rotation curve
interior and exterior to the solar circle and the mass-to-light ($M/L$)
ratios of the disk and spheroid.
The shape of dark matter halos is still an open question (for a review, see
Ashman 1992), but substantially flattened halos (E4-E6) have been
indicated in some studies (Dubinski \& Carlberg 1991; Katz 1991; Katz \&
Gunn 1991; and Sackett \& Sparke 1990.)

Since we do not know the exact Galactic rotation curve,
the shape of the Galactic halo, or precise
disk and bulge normalizations,
we choose arbitrary values for these quantities within
the range allowed by observations.  We then find the best fit values
of $v_\infty$ and $a$ for a spherical halo and for an E6 (axis ratio
$c/a = 0.4$ in density).  For each of these halo models, we determine
the optical depth to microlensing by \ms\
toward the LMC, the SMC, and the Galactic bulge (choosing
Baade's Window to be specific).

Although the optical depth toward individual lines of sight can vary
by a factor of a few between models, we find that the ratio
of optical depth toward the SMC over that to the LMC, \tsmctlmc, varies by
only a few percent over all Galactic parameters except halo
flattening, to which it is much more sensitive.
For a spherical halo, the ratio is \tsmctlmc\ $\sim 1.45$;
for an E6 halo, \tsmctlmc\ $\sim 0.95$.
If the halo is not truncated, these ratios vary by only
about 3-5\%
over the whole range of models considered, while
the individual optical depths to the LMC and SMC
vary by as much as a factor of 3 to 4 between models.
For a truncated halo, the \tsmctlmc\ ratios
vary a bit more, by as much as 8\%.
For the most extreme case that we examined, a highly flattened (E6)
and closely truncated halo ($2R_0$), the optical depth toward the LMC
decreases by a factor of $\sim 2$
relative to a standard model.  For
such a halo, the current searches are sensitive only to \ms\ of
mass  $M\, \lsim \, {\cal O}(10)\,M_\odot$.  By comparison,
if one assumes a spherical untruncated halo, then a null result would rule
out \ms\ to as high as
$M \, \lsim \, {\cal O}(10^6)\,M_\odot$ (Gould 1992).

In contrast to the optical depths toward the LMC and SMC,
the optical depth toward the bulge seems to provide very little
model-independent information.  The principal value of the bulge
observations to constraining the mass distribution of the Galaxy
is that they will allow us to search for dark matter in the disk.

The remainder of the paper is organized as follows:  In \S\ 2,
we derive expressions for the optical depth
along arbitrary lines of sight to microlensing from within
an isothermal halo of arbitrary flattening,
and discuss its dependence on halo parameters and the geometry of the
line of sight relative to the halo.
In \S\ 3, we describe our procedure for determining
the best fit halo parameters
for various observationally constrained models of the
Galactic rotation curve, disk, and bulge.
In \S\ 4 we compare, for a wide range of Galactic parameters,
the optical depths of E0 versus E6 untruncated halos
along lines of sight to the LMC, SMC, and the Galactic bulge;
in \S\ 5 a similar comparison is made for truncated halos.
We argue from these results that if the Galactic dark halo is composed
of \ms, the ratio \tsmctlmc\ is an reliable measure of the ellipticity
of the halo.
In \S\ 6, we discuss limitations imposed by statistical fluctuations
and various lensing ``backgrounds.''
We summarize our conclusions in \S\ 7,
and in the Appendix we invert the $R^{1/4}$ law to derive analytic
expressions for the volume density and corresponding force from a
spherical de~Vaucouleurs bulge.

\chapter{Optical Depth to Microlensing by a Flattened Halo}

To model the Galactic dark halo, we
construct a four-parameter family of flattened
density profiles $\rho(\br)$
\foot{We note that this formulation for
the halo truncation radius is not necessarily physical since it does
not represent a solution to the collisionless Boltzmann equation.  The
optical depth to microlensing, however, is not sensitive to the
precise shape of the truncation profile and, as we shall see, is only
weakly dependent on the location of even a rather severe truncation.
Similarly, we do not consider here the effect of anisotropic orbits of
\ms\ in a flattened halo.  The degree of anisotropy will depend on the
amount of rotation support in the halo.  Although any anisotropy in
the orbits could alter estimates of the timescales of lensing events
by as much as factor of two, the primary conclusions of this paper,
which involve optical depths and their ratios, would not be
affected.
},
$$\rho(\br) = {\tan\psi \over \psi}
{v_\infty^2\over 4\pi G}\, \biggl( {1\over a^2 + \zeta^2} \biggr) \,
\theta(R_T - \zeta),\eqn\rhoofrp$$
where
$$\zeta^2 \equiv r^2 + z^2\tan^2\psi,\eqn\oblatedef$$
and where $r$ is the Galactocentric radius and $z$ is the
height above the Galactic plane.
The flattening parameter,
$\psi$, is defined such that $\cos\psi$ is the axis ratio $q = c/a$
of the halo density profile
(which is assumed to be axisymmetric in the $z=0$ plane).
The remaining three parameters, $v_\infty$, $a$, and
$R_T$, in equation \rhoofrp\ are defined as in equation
\rhoofr.
Note that the halo is truncated on an isodensity contour
that crosses the Galactic plane at $R_T$; since the halo is assumed to be
oblate, the truncation distance in other directions is smaller.

A \m\ of mass $M$
in this halo will have an Einstein ring radius, $r_*$, given by
(\eg, Blandford \& Kochanek 1987)
$$r_*^2 = {4 G M L (D-L)\over c^2\,D},\eqn\reeq$$
where $D$ is the distance from the observer to the source
and $L$ is the distance from the observer to the lensing MACHO.
The
cross section to lensing
is defined by convention to be
the probability that a given observed source lies within
an angle $r_*/L$ of a \m.
If the halo is composed
of MACHOs, the optical depth along a line of sight
parameterized by $L$, is therefore given by
$$\tau = \int_0^D dL {4\pi G\rho[\br(L)]\over c^2}L
\biggl(1-{L\over D}\biggr).
\eqn\tauone$$

Substituting equation \rhoofrp\ into equation \tauone\
leads to the optical depth toward Galactic coordinates $\ell$ and $b$
$$\tau(\ell, b) = {\tan\psi \over \psi} {v_\infty^2 \over c^2} {1 \over D}
\int_0^{D_T} { {dL~~~(D - L) L} \over { (a^2 + R_0^2) -
(2 R_0 \cos\ell \cos b)\,
L + (1 + \sin^2 b \tan^2 \psi)\,
L^2} }~ ,
\eqn\tauint$$
where
$R_0$ is the Galactocentric radius at the solar circle and $D_T$ is
the distance to the source or the truncation distance along the
line of sight,
whichever is smaller.  By completing
the square in the denominator, this integral can be evaluated to yield
$$\tau(\ell,b) = {\tan\psi \over \psi}{v_\infty^2\over {c^2 p^2}}
{1\over D}\biggl[-s + \biggl({D\over 2} - \Delta\biggr)\ln(s^2+q^2)
+ \biggl({q^2 + D\Delta - \Delta^2\over q}\biggr)\arctan{s\over q}
\biggr]_{s=-\Delta}^{s=s_{\rm max}}\eqn\taucomp$$
where
$$ p^2 \equiv 1 + \tan^2\psi\sin^2 b;\qquad \Delta\equiv {R_0\cos\ell\cos b
\over p^2}~, \eqn\panddel$$
$$ q^2 \equiv {R_0^2 + a^2\over p^2} - \Delta^2,~~~{\rm and}\eqn\qdef$$
$$ s_{\rm max} \equiv {\rm min}\biggl(D-\Delta,\sqrt{{R_T^2-R_0^2\over p^2}
+\Delta^2}\biggr)~. \eqn\sdef$$

Once the parameters of a given model are established, it is straightforward
to evaluate the optical depth in any direction using equation \tauint.

One can see from equation \tauint\ that information about the shape of the dark
halo is carried in the prefactor $(\tan \psi/\psi)$ and in the factor
$(1 + \sin^2 b \tan^2 \psi)$ that
multiplies $L^2$ in the denominator of the integrand.  The prefactor,
which decouples from other halo parameters and from the source direction, is
equal to unity for a spherical halo, and 1.98 for an E6 halo.  This is
the only dependence of $\tau$ on the flattening of the halo for lines of
sight that lie in the Galactic plane.  In principle, it might seem that
this would imply that searches for \ms\ toward the bulge would provide
a strong discriminator for the shape of the dark halo, since the
optical depth would be increased by a factor of two for a substantially
flattened (E6) halo.
However, as we shall discuss in \S\ 3, in practice the poorly known
halo parameters, $a$ and $v_\infty$,
and microlensing contamination from disk stars render the bulge fields
nearly useless as probes of the ellipticity of the dark halo.

High latitude sources are more effective at measuring halo shape because
of decreased disk contamination and because of the $L^2$ term
in the denominator of the integrand of equation \tauint.
Shown in Figure 1 are the optical depths toward
SMC ($\ell=303\deg, b=-44\deg$) and
LMC ($\ell=280\deg, b=-33\deg$) fields for both E0 and E6 halos as a
function of halo core radius.  A somewhat arbitrary, but reasonable,
halo asymptotic speed of $v_\infty = 220 \kms$ was used for the purposes of
the calculation, but since from equation \tauint\
it is clear that $\tau$ varies strictly as the square of $v_\infty$,
these results can be rescaled to any other value.  The decline of optical
depths with halo core radius is easily understood from Eq. \tauint;
the decline with halo truncation
can be a factor of two in the most severe case.
In general, \tsmc\ is larger tha \tlmc\ because the line of sight to the
SMC passes closer to the center of the halo;
$\cos\ell \cos b = 0.39$ for the SMC and 0.15 for the LMC.
The optical depth to the SMC is more sensitive
to the shape of the halo because of its larger Galactic latitude.  Note
that along the LMC line of sight, the effect of the $(\sin^2b \tan^2\psi)$ term
in the denominator is nearly offset by the prefactor $(\tan\psi/\psi)$ in
the numerator for an E6 halo; toward the SMC,
the $(\sin^2b \tan^2\psi)$ term is substantially larger,
and thus the gap between E0 and E6 curves in Figure 1 is wider
for the SMC than the LMC.

Difficulty in determining the core radius and asymptotic speed of the
halo translate into substantial uncertainties in the individual opticals
depths toward the SMC and LMC.  In the {\it ratio\/} of
optical depths \tsmctlmc, however, the asymptotic halo speed, $v_\infty$, drops
out completely, and, as can be seen from Figure 2, the remaining dependence
on the core radius, $a$, is a rather weak one.  Furthermore,
regardless of the truncation radius of the halo and the value of $a$
(up to more than 20 \kpc), \tsmctlmc\ is 20 - 40\% lower for an E6 than an
E0 halo.  We thus propose \tsmctlmc\ as an indicator of the ellipticity of
the dark halo.  What remains to be shown is that the range of core radii
shown in Figures 1 and 2 are appropriate to the Milky Way, and that this
difference is one that can be detected in on-going \m\ searches; this we
do in subsequent sections.

\chapter{Description of Models}

We model the mass of the Galaxy in terms of three components,
a disk, a spheroid, and a dark halo.
In order to use the Galactic rotation curve as a constraint on the dark
halo parameters $v_\infty$ and $a$ for a halo of given flattening, we must
evaluate the contribution of the disk and spheroid to the rotational
support as a function of radius.

The disk we take as
a double exponential in density, $\rho(R,z) = \rho_0 \ e^{-|z|/z_0} \
e^{-R/h}$, with scale height, $z_0=350\,$pc and
scale length, $h=3.5\,$kpc (Binney \& Tremaine 1987).
The disk is then completely specified by a
single parameter, the local column density, $\Sigma_0$,
which is related to the central volume density, $\rho_0$, of the disk through
$\Sigma_0 = 2 \ z_0 \ \rho_0 \ e^{-R_0/h}$.

Bahcall (1984) estimates the total mass of
the visible components of the disk in the local neighborhood
to be $\Sigma_0\sim 50\,M_\odot\,\rm pc^{-2}$,
which places a lower limit on the surface mass density of the Galactic disk.
The disk may also have dark components (Bahcall, Flynn, \& Gould 1992);
a so-called ``maximal disk'' (which is often used to model the
mass distribution in other galaxies) provides a hard upper limit on
$\Sigma_0$ of about twice that seen in visible components.
The total column density of the disk is further
constrained by the measurement
of Kuijken \& Gilmore (1989, KG), but not as tightly as those authors
suggest.  KG show a best fit to their data (prior to
imposing the rotation-curve constraint) of $\Sigma_0=71\pm 6\, M_\odot\,
\rm pc^{-2}$ and essentially no dark halo.  After imposing the rotation
constraint, they find a best fit of $\Sigma_0=46\, M_\odot\,\rm pc^{-2}$.
However, Gould (1990) showed that this estimate was statistically biased
and that a better treatment yields $\Sigma_0=54\pm 9\, M_\odot\,\rm pc^{-2}$.
Moreover, the best combined fit of the data and the rotation curve
constraint is not
particularly good, so that the confidence of the rotation constraint
may have been overestimated.  Hence, it seems conservative to allow a local
column density as high as $\Sigma_0=75\,M_\odot\,\rm pc^{-2}$.
In our models, we examined
$\Sigma_0=50$ and $75\,M_\odot\,\rm pc^{-2}$.
\foot{
Although maximal disk models are widely and, in some respects, successfully
used in fitting the rotation curves of external galaxies, we do not
include here models
with a Galactic maximal disk of $\Sigma_0=100\,M_\odot\,\rm pc^{-2}$
because such models are inconsistent with the data of Kuijken \& Gilmore
(1989).
In any event, the
amount of disk dark matter will be directly measured by the \m\ and \wc\
experiments themselves (Paczy\'nski 1991; Griest \etal 1991).
}

The other major visible component of the Galaxy is the spheroid,
which has a total luminosity of $2.4\times 10^9\,L_{\odot,{\rm V}}$
(Bahcall \& Soneira 1980),
after correction to the current IAU standard of $R_0 = 8.5$ kpc
for the solar Galactocentric distance.
The total mass in the spheroid is uncertain because its
mass-to-light ratio is not known.  The spheroids of spiral galaxies are
often thought to be similar to elliptical galaxies and
the mass-to-light ratios of the
inner parts of elliptical galaxies is estimated to be $\sim 10\,M_\odot/
L_{\odot,{\rm V}}$ (\cf Lauer 1985; Peletier 1989).
On the other hand, the mass-to-light ratios of the stellar
disks of spirals are typically $\lsim 5\,M_\odot/L_{\odot,{\rm V}}$
(\cf Begeman 1987; Kent 1987).
We therefore consider models with total spheroid masses of
$1.2\times 10^{10}M_\odot$ and $2.4\times 10^{10}M_\odot$.
For the density profile of the spheroid, we adopt
the spherical distribution that produces
a de~Vaucouleurs $R^{1/4}$ law
in projection.
(A derivation of this volume density profile and the resulting expression for
the gravitational acceleration it provides can be found in the Appendix.)
We took the effective radius, $R_e$, of the spheroid to be 2.87 kpc,
which agrees with that from Bahcall and Soneira after rescaling to
$R_0 = 8.5$ kpc.

The third element in our model of the Galaxy is the rotation
curve.  The Galactic rotation curve is approximately flat between 3 and 17 kpc,
but is consistent with rising or falling by 14\%
(or 30 \kms) over that range (Fich, Blitz \& Stark 1989).
Although the Galactic rotation curve is consistent with being flat
where it has been measured, the slope is tied to local values of $R$ and $v$
(Schechter 1993).  This means that because the the rotation curve is generally
measured in fundamentally different ways
inside and outside the solar circle, it is possible that the
rotation curve rises inside the solar circle, but falls outside it,
or vice versa.
We were therefore led to examine a total of nine model rotation curves
that span this range of possibilities.
Each curve is pinned at the solar circle to have the
IAU value of
$220 \kms$ for the circular speed of the local standard
of rest.
This value has become somewhat controversial
of late (for a discussion, see Merrifield 1992; Fich \& Tremaine 1991).
However, the optical depth of a halo having a smaller rotation
speed than the standard value by $\eta$ can be found
by incorporating a disk and spheroid that are larger by $\eta^{-2}$
and scaling the resulting optical depth down by $\eta^2$.
The speeds at 3 and 17~kpc are allowed to vary up or down from this
local speed and a linear extrapolation is taken to the local neighborhood;
the total deviation from flatness is less than or equal to
a 14\% difference in rotation speed in this range.
For example, the model rotation
curve labeled ``rise-rise'' in our set of nine corresponds to a linear
rise in rotation speed from 208.2 to $220 \kms$ from 3 to 8.5 kpc, and then
a linear rise to $238.2 \kms$ to 17 kpc.

Once the disk, spheroid, and rotation curve are specified, we find
the best fit spherical halo by setting the flattening parameter
$\cos\psi=1$ in equation \rhoofrp, and minimizing $\chi^2$ with
respect to $v_\infty$ and $a$.  We then
repeat the procedure for an E6 halo by setting $\cos\psi=0.4$.
The optical depths along various lines of sight are then calculated
using equation \taucomp, and assuming distances to the LMC and SMC of 50 and
63 kpc respectively (Binney \& Tremaine 1987).

Within the constraints provided by observations, our choice for the
parameterization of the Galactic rotation curve is arbitrary.  In many
cases the minima in $\chi^2$ were quite broad, so that the halo core
radius, especially, was poorly constrained.
Since the fits were not made to data points weighted by observational
uncertainties, but to the model curves,
the $\chi^2$ values associated with the fits are not meaningful for
testing the absolute goodness-of-fit for a particular model, or the relative
goodness-of-fit between models.
We stress that our only aim was to explore a reasonable
range of the halo parameters $a$ and $v_\infty$, constrained by
our knowledge of the rotation curve of the Galaxy and the mass of its visible
components.  As we will show, the most important diagnostic for halo
flattening is independent of the present uncertainty in these quantities.

\chapter{Untruncated Halos}

Our results for untruncated halos (that is, halos that extend to at least
the Magellanic clouds) are illustrated in Table 1, where we display
optical depths and ratios of optical depths for
Galactic models consisting of a disk with no dark mass normalized locally
at $\Sigma_0=50\,M_\odot\,\rm pc^{-2}$ and a relatively light
spheroid with a mass-to-light ratio of $5 M_\odot/L_{\odot,{\rm V}}$.  The
nine columns of the table represent nine different rotation curves
that span current observational
constraints, as described above.
The first two rows give the core radius, $a$ (in kpc), of the best fit
models for a spherical and an E6 halo; rows 3 and 4
show the corresponding asymptotic circular speed, $v_\infty$, of the halo.
Rows 5, 6, and 7
show the optical depth toward the LMC, the SMC, and Baade's
Window for a spherical halo.  Row 8 gives the ratio of the optical
depth toward the LMC for an E6 relative to a spherical halo.
Rows 9 and 10 indicate the ratio of the optical depth toward the SMC
compared to that toward the LMC for a spherical and E6 halo, respectively.
Rows 11 and 12
give the ratios of the optical depth to Baade's Window compared
to the LMC for spherical and E6 halos.

\singlespace
\midinsert
\def\lmc{{\rm LMC}}
\def\smc{{\rm SMC}}
\def\bw{{\rm BW}}
\def\kms{{\rm km}\ {\rm s}^{-1}}
\def\kpc{{\rm kpc}}
\def\Ez{{\rm (E0)}}
\def\Es{{\rm (E6)}}
\def\llap#1{\hbox to 6.3in{\hss#1}}
\llap{
$$\vbox{\halign{#\hfil\tabskip=1em plus1em minus1em&\hfil#&\hfil#&\hfil#
&\hfil#&\hfil#&\hfil#&\hfil#&\hfil#&\hfil#\cr
\multispan{10}{\hfil TABLE 1\hfil}\cr
\noalign{\medskip}
\multispan{10}{\hfil Optical Depths for Micro-Lensing (Light Disk
and Spheroid)\hfil}\cr
\multispan{10}{\hfil $M_{\rm sph} = 1.2 \times 10^{10} M_\solar$\ \ \ \
$\Sigma_0 = 50~M_\solar~{\rm pc}^{-2}$\hfil}\cr
\noalign{\medskip\hrule\smallskip\hrule\medskip}
&Fall-\hfill&Fall-\hfill&Fall-\hfill&Flat-\hfill&Flat-\hfill
&Flat-\hfill&Rise-\hfill&Rise-\hfill&Rise-\hfill\cr
Rotation Curve&Fall\hfill&Flat\hfill&Rise\hfill&Fall\hfill
&Flat\hfill&Rise\hfill&Fall\hfill&Flat\hfill
&Rise\hfill\cr
\noalign{\medskip\hrule\medskip}
$a(\kpc) \ \Ez $ &     0.00 &     0.00 &     0.33 &     0.00 &     0.24 &
0.86 &     0.16 &     0.79 &     1.59 \cr
$a(\kpc) \ \Es $ &     0.00 &     0.00 &     0.41 &     0.00 &     0.29 &
0.96 &     0.20 &     0.94 &     1.70 \cr
$v_c(\kms) \ \Ez $ &   162 &   170 &   184 &   158 &   170 &   190 &   158 &
176 &   200 \cr
$v_c(\kms) \ \Es $ &   162 &   170 &   183 &   158 &   170 &   188 &   157 &
176 &   196 \cr
$\tau_{\lmc} \times 10^6 \ \Ez $ &     0.339 &     0.373 &     0.436 &
0.325 &     0.375 &     0.463 &     0.321 &     0.401 &     0.508 \cr
$\tau_{\smc} \times 10^6 \ \Ez $ &     0.499 &     0.549 &     0.642 &
0.479 &     0.553 &     0.682 &     0.473 &     0.590 &     0.747 \cr
$\tau_{\bw} \times 10^6 \ \Ez $ &     0.527 &     0.580 &     0.631 &     0.505
&     0.562 &     0.521 &     0.490 &     0.467 &     0.404 \cr
$\tau_{\lmc} \ ({\rm E6/E0}) $ &     1.002 &     1.002 &     0.999 &     1.002
&     1.001 &     0.988 &     1.002 &     0.999 &     0.970 \cr
$\tau_{\smc}/\tau_{\lmc} \ \Ez $ &     1.474 &     1.474 &     1.473 &
1.474 &     1.473 &     1.472 &     1.474 &     1.472 &     1.469 \cr
$\tau_{\smc}/\tau_{\lmc} \ \Es $ &     0.962 &     0.962 &     0.962 &
0.962 &     0.962 &     0.962 &     0.962 &     0.962 &     0.962 \cr
$\tau_{\bw}/\tau_{\lmc} \ \Ez $ &     1.556 &     1.556 &     1.448 &     1.556
&     1.497 &     1.124 &     1.527 &     1.166 &     0.796 \cr
$\tau_{\bw}/\tau_{\lmc} \ \Es $ &     1.915 &     1.916 &     1.858 &     1.915
&     1.886 &     1.647 &     1.901 &     1.654 &     1.313 \cr
\noalign{\medskip\hrule}
}}$$
}
\endinsert
\normalspace

Across the range of rotation curves considered, the best fit core radii and
asymptotic circular halo speeds displayed in Table~1 vary
significantly --- from about 0 to 1.7 kpc, and 160 to
$200\,\kms$, respectively --- but these quantities
vary much less between the E0 and E6 models for a given model rotation curve.
This is especially
true of asymptotic circular speed.  The optical depth toward
each of the three lines of sight varies by
about a factor of 1.5 across the range of model rotation
curves.
Nevertheless,
the {\it ratio\/} of the optical depths toward the SMC and the LMC is constant
over the entire range of models to within 1\%.  This is true separately
for the E0 and E6 halos, but the ratio is different in the two
cases.  For E0, \tsmctlmc\ $=1.47$, while for E6,
\tsmctlmc\ $=0.96$.
As anticipated in \S\ 2,
this indicates that for the range of models considered in Table 1,
the ratio \tsmctlmc\ is an excellent indicator of halo flatness.
By contrast, the ratio toward Baade's window
in the Galactic bulge,
$\tau_{\scriptscriptstyle \bw}/\tau_{\scriptscriptstyle \lmc}\,$,
varies by more than
a factor of nearly 2.  Finally, note from row
8 that \tlmc\ is almost exactly the same for an E0 as an E6 halo
for each rotation curve considered.

In Table~2 we present results for a model in which the observed components
of the Galaxy are substantially heavier.  The disk is increased in mass
by 50\% and the spheroid by 100\%.

\singlespace
\midinsert
\def\lmc{{\rm LMC}}
\def\smc{{\rm SMC}}
\def\bw{{\rm BW}}
\def\kms{{\rm km}\ {\rm s}^{-1}}
\def\kpc{{\rm kpc}}
\def\Ez{{\rm (E0)}}
\def\Es{{\rm (E6)}}
\def\llap#1{\hbox to 6.3in{\hss#1}}
\llap{
$$\vbox{\halign{#\hfil\tabskip=1em plus1em minus1em&\hfil#&\hfil#&\hfil#
&\hfil#&\hfil#&\hfil#&\hfil#&\hfil#&\hfil#\cr
\multispan{10}{\hfil TABLE 2\hfil}\cr
\noalign{\medskip}
\multispan{10}{\hfil Optical Depths for Micro-Lensing (Heavy Disk and
Spheroid)\hfil}\cr
\multispan{10}{\hfil $M_{\rm sph} = 2.4 \times 10^{10} M_\solar$\ \ \ \
$\Sigma_0 = 75~M_\solar~{\rm pc}^{-2}$\hfil}\cr
\noalign{\medskip\hrule\smallskip\hrule\medskip}
&Fall-\hfill&Fall-\hfill&Fall-\hfill&Flat-\hfill&Flat-\hfill
&Flat-\hfill&Rise-\hfill&Rise-\hfill&Rise-\hfill\cr
Rotation Curve&Fall\hfill&Flat\hfill&Rise\hfill&Fall\hfill
&Flat\hfill&Rise\hfill&Fall\hfill&Flat\hfill
&Rise\hfill\cr
\noalign{\medskip\hrule\medskip}
$a(\kpc) \ \Ez $ &     0.00 &     0.11 &     1.77 &     0.00 &     1.58 &
6.39 &     1.48 &     5.17 &    11.6 \cr
$a(\kpc) \ \Es $ &     0.00 &     0.13 &     1.81 &     0.00 &     1.56 &
6.76 &     1.74 &     6.01 &    12.6 \cr
$v_c(\kms) \ \Ez $ &   117 &   130 &   163 &   113 &   144 &   224 &   124 &
184 &   300 \cr
$v_c(\kms) \ \Es $ &   117 &   130 &   160 &   113 &   140 &   216 &   124 &
184 &   292 \cr
$\tau_{\lmc} \times 10^6 \ \Ez $ &     0.178 &     0.217 &     0.339 &
0.164 &     0.264 &     0.532 &     0.197 &     0.381 &     0.710 \cr
$\tau_{\smc} \times 10^6 \ \Ez $ &     0.263 &     0.320 &     0.498 &
0.242 &     0.387 &     0.760 &     0.289 &     0.549 &     0.997 \cr
$\tau_{\bw} \times 10^6 \ \Ez $ &     0.277 &     0.335 &     0.250 &     0.256
&     0.211 &     0.114 &     0.165 &     0.104 &     0.078 \cr
$\tau_{\lmc} \ ({\rm E6/E0}) $ &      1.002 &     1.000 &     0.960 &     1.002
&     0.956 &     0.961 &     0.997 &     1.001 &     1.015 \cr
$\tau_{\smc}/\tau_{\lmc} \ \Ez $ &     1.474 &     1.474 &     1.468 &
1.474 &     1.469 &     1.430 &     1.470 &     1.440 &     1.404 \cr
$\tau_{\smc}/\tau_{\lmc} \ \Es $ &     0.962 &     0.962 &     0.963 &
0.962 &     0.963 &     0.971 &     0.963 &     0.969 &     0.996 \cr
$\tau_{\bw}/\tau_{\lmc} \ \Ez $ &     1.556 &     1.541 &     0.739 &     1.556
&     0.801 &     0.215 &     0.837 &     0.272 &     0.109 \cr
$\tau_{\bw}/\tau_{\lmc} \ \Es $ &     1.915 &     1.909 &     1.271 &     1.915
&     1.374 &     0.374 &     1.299 &     0.430 &     0.174 \cr
\noalign{\medskip\hrule}
}}$$
}
\endinsert
\normalspace

General comments made about Table 1 also apply to Table 2,
except that the core radii and asymptotic circular speeds
vary over a wider range (now, $0 \le a \le 12$ kpc and
$110 \le v_\infty \le 300$ \kms).
The full range in halo core radius and asymptotic speeds spanned by our
models agrees favorably with that inferred for external Sb and Sc spirals
($2.5 \le a \le 15$ kpc and $116 \le v_\infty \le 307$ \kms, Begeman 1987),
and with the values of $a = 3$ kpc and $v_\infty = 230$ \kms advocated
by Merrifield (1992) for the Milky Way.
Note that the range of best-fit halo core radii lie comfortably within
the range plotted in Figures 1 and 2.  We should not be surprised, therefore,
that
the ratio of \tlmc\ for E6 compared to that for E0 halos (row 8), and the
ratio \tsmctlmc\ for both E0 (row 9) and E6 (row 10) halos
are nearly constant across the row,
\ie they are nearly independent of the assumptions made about the
shape of the Galactic rotation curve between 3 and 17 kpc.
Not only are the values constant within
a row, but they are equal to the values in the corresponding row in
Table 1.  That is, we find that for untruncated halos,
\tsmctlmc\ depends only on the ellipticity of the halo and
not on the details of the rotation curve or on the masses of the disk
and the spheroid.
Furthermore, for any of mass models that we explored that led
to reasonable values of the halo parameters, we find that
\tlmc\ (E6)/\tlmc\ (E0) $\sim 1$, that is: for untruncated halos,
the optical depth
toward the LMC is independent of the halo flattening regardless of the
value of other Galactic parameters.

\chapter{Truncated Halos}

By introducing a truncation radius, $R_T$, for the halo, one
introduces additional uncertainty into the problem; we describe here
our motivation for considering this possibility.  From the motion
of distant globular clusters and satellite galaxies, it is believed
that the dark halo must extend at least to
$\sim 3 R_0$ (for a review, see Ashman 1992 and reference therein).
Zaritsky \etal (1992) have shown that the dark halo
in other spiral galaxies extends to several hundred kpc.
Thus, if the dark halo of the Milky Way is similar to that of other
galaxies, it might seem unnecessary to explore the effect of halo truncation.
\m\ searches, however, are not sensitive
to dark matter {\it per se}, but only to the fraction of the dark
matter
in compact objects.  It is quite possible there are two forms
of dark matter, baryonic and non-baryonic.  The baryonic
dark matter, being dissipational, would form \ms\ that might
lie in a relatively compact (and perhaps flattened) halo, while the
non-baryonic matter might form a more extended halo.  The
observed orbits of satellite galaxies and other dynamical arguments
would not be sensitive to such a distinction.  The introduction
of a truncation radius for \ms, therefore, allows us to explore
the possibility that {\it baryonic\/} dark matter does not extend very far
beyond the luminous disk of our Galaxy.

To examine this question, we computed optical depths along the same
lines of sight for the same Galactic parameters used for untruncated halos,
but truncated the dark halo on isodensity surfaces that cross the Galactic
plane at twice or four times the solar circle.
In Table 3, we compare truncated and untruncated halos for
the class of models considered in Table 1, \ie those with
$\Sigma_0=50\,M_\odot\,\rm pc^{-2}$ and $M_{\rm sph}=1.2\times 10^{10}M_\odot$.
The subscripts `2' and `4'
refer to halo truncation
at $R_T=2 R_0$ and $R_T=4 R_0$ respectively.  The subscript `$\infty$'
denotes an untruncated halo.  The first four rows give the
ratios of the optical depths of truncated to untruncated halos toward
the LMC and the SMC for spherical halos of various truncation radii.
Rows 5, 6, and 7 give the
ratio of the optical depths toward the LMC in E6 compared to E0 halos.
Rows 8, 9, and 10 give the ratios of the optical depths toward the
SMC compared to the LMC for spherical halos of
differing $R_T$; the final three rows give the same quantities
for E6 halos.

\singlespace
\midinsert
\def\lmc{{\rm LMC}}
\def\smc{{\rm SMC}}
\def\bw{{\rm BW}}
\def\kms{{\rm km}\ {\rm s}^{-1}}
\def\kpc{{\rm kpc}}
\def\Ez{{\rm (E0)}}
\def\Es{{\rm (E6)}}
\def\llap#1{\hbox to 6.5in{\hss#1}}
\llap{
$$\vbox{\halign{#\hfil\tabskip=1em plus1em minus1em&\hfil#&\hfil#&\hfil#
&\hfil#&\hfil#&\hfil#&\hfil#&\hfil#&\hfil#\cr
\multispan{10}{\hfil TABLE 3\hfil}\cr
\noalign{\medskip}
\multispan{10}{\hfil Optical Depths for Micro-Lensing by Truncated Halos (Light
Disk and Spheroid)\hfil}\cr
\multispan{10}{\hfil $M_{\rm sph} = 1.2 \times 10^{10} M_\solar$\ \ \ \
$\Sigma_0 = 50~M_\solar~{\rm pc}^{-2}$\hfil}\cr
\noalign{\medskip\hrule\smallskip\hrule\medskip}
&Fall-\hfill&Fall-\hfill&Fall-\hfill&Flat-\hfill&Flat-\hfill
&Flat-\hfill&Rise-\hfill&Rise-\hfill&Rise-\hfill\cr
Rotation Curve&Fall\hfill&Flat\hfill&Rise\hfill&Fall\hfill
&Flat\hfill&Rise\hfill&Fall\hfill&Flat\hfill
&Rise\hfill\cr
\noalign{\medskip\hrule\medskip}
$\tau_{\lmc,4}/\tau_{\lmc,\infty} \ \Ez $ &     0.944 &     0.944 &     0.944 &
    0.945 &     0.944 &     0.934 &     0.944 &     0.937 &     0.917 \cr
$\tau_{\lmc,2}/\tau_{\lmc,\infty} \ \Ez $ &     0.619 &     0.619 &     0.614 &
    0.619 &     0.615 &     0.567 &     0.615 &     0.583 &     0.495 \cr
$\tau_{\smc,4}/\tau_{\smc,\infty} \ \Ez $ &   0.917 &     0.917 &     0.915 &
  0.917 &     0.916 &     0.898 &     0.916 &     0.904 &     0.868 \cr
$\tau_{\smc,2}/\tau_{\smc,\infty} \ \Ez $ &   0.646 &     0.646 &     0.640 &
  0.646 &     0.642 &     0.583 &     0.642 &     0.602 &     0.499 \cr
$\tau_{\lmc,\infty}\ ({\rm E6/E0}) $ &     1.002 &     1.000 &     0.960 &
1.002 &     0.956 &     0.961 &     0.997 &     1.001 &     1.015 \cr
$\tau_{\lmc,4}\ ({\rm E6/E0}) $ &     0.861 &     0.859 &     0.823 &     0.861
&     0.820 &     0.805 &     0.855 &     0.841 &     0.812 \cr
$\tau_{\lmc,2}\ ({\rm E6/E0}) $ &     0.770 &     0.768 &     0.734 &     0.770
&     0.733 &     0.706 &     0.762 &     0.733 &     0.689 \cr
$\tau_{\smc,\infty}/\tau_{\lmc,\infty} \ \Ez $ &     1.474 &     1.474 &
1.468 &     1.474 &     1.469 &     1.430 &     1.470 &     1.440 &     1.404
\cr
$\tau_{\smc,4}/\tau_{\lmc,4} \ \Ez $ &     1.431 &     1.431 &     1.424 &
1.430 &     1.425 &     1.374 &     1.426 &     1.389 &     1.329 \cr
$\tau_{\smc,2}/\tau_{\lmc,2} \ \Ez $ &     1.539 &     1.538 &     1.531 &
1.539 &     1.533 &     1.471 &     1.533 &     1.488 &     1.415 \cr
$\tau_{\smc,\infty}/\tau_{\lmc,\infty} \ \Es $ &     0.962 &     0.962 &
0.963 &     0.962 &     0.963 &     0.971 &     0.963 &     0.969 &     0.996
\cr
$\tau_{\smc,4}/\tau_{\lmc,4} \ \Es $ &     0.867 &     0.867 &     0.866 &
0.867 &     0.866 &     0.857 &     0.865 &     0.859 &     0.851 \cr
$\tau_{\smc,2}/\tau_{\lmc,2} \ \Es $ &     0.886 &     0.885 &     0.884 &
0.885 &     0.884 &     0.872 &     0.884 &     0.874 &     0.861 \cr
\noalign{\medskip\hrule}
}}$$
}
\endinsert
\normalspace

The most remarkable thing about Table 3 is that all 27 of the entries
for the ratio \tsmctlmc\ for truncated, spherical halos
(rows 8, 9, and 10) and for truncated, flattened (E6) halos (rows 11, 12,
and 13) are approximately equal.
We found this to be true for all models that we explored that
gave reasonable best-fit halo parameters.

In Figure 3, we present the ratio \tsmctlmc\ as a function of
the assumed Galactic rotation curve model for each of the four mass models
(four combinations of light and heavy disk and bulge) at each of three
halo truncations.  It is clear from this figure that the optical depth
ratio \tsmctlmc\ is quite insensitive to assumptions about the rotation
curve and $M/L$ of the luminous components of the Galaxy.  The ratio is
sensitive to the halo truncation radius, but even allowing for the possibility
of strong truncation (at $2 R_0$), the difference in \tsmctlmc\ for E6 as
opposed to E0 is unambiguous, and as we shall show in the next section,
should be measurable.  {\it We conclude that,
the ratio of optical depths, \tsmctlmc, is a robust,
model-independent indicator of halo flatness.\/}

\chapter{Detectability}

If \ms\ are
not detected by ongoing experiments, how would allowance for
ellipticity and truncation of the halo affect the interpretation
of this null result?  If \ms\ are detected, to what accuracy
can the ellipticity and truncation of the halo be measured?
In order to address these questions, we must review the current \m\ search
technique.

The characteristic time scale of a \m\ event is
the time taken by the \m\ to cross the Einstein ring radius:
$$\omega^{-1}\equiv {r_*\over v_t} =
{\sqrt{4 G M L (1-L/D)}\over v_t c}
\sim 70\,{\rm days}\,(M/M_\odot)^{1/2},\eqn\machotime$$
where $v_t$ is the transverse speed and the evaluation is for ``typical''
parameters of $L\sim 10\,$kpc, $v_t\sim 200\,\kms$, and $L/D \sim 1/5$.
The expected number
of events underway at any given time is $N_*\tau$, where $N_*$
is the number of stars being observed and $\tau$ is the optical depth.
Note that if $M\,\gsim\,100\,M_\odot$, then the characteristic time of a
\m\ event becomes several years, which is comparable to
or longer than the
planned four years of the \m\ Collaboration experiment.
In this case, the full
light curve would not be observed.  For this reason, it was originally
believed that the experiment would be insensitive to \ms\ in this high
mass range.  However, Gould (1992) showed that candidate events could
be recognized from the data for \ms\ as massive as $10^6\, M_\odot$ and
that these candidates could be distinguished from backgrounds (such as
variable stars) by a variety of techniques.  For example, for masses
$M\,\gsim\,10^3 M_\odot$, the two lensed stellar images could be resolved by
(an unrepaired) Hubble Space Telescope, and further confirmation could be
obtained by measuring the proper motion of these images.

The total number of
observed
events depends on the characteristic mass
of the \ms.  If the events are longer than the
duration of the experiment, then the approximately
$N_*\tau$ events that are taking place on the first day of the
observations will be the only events during the entire span of observations.
If the events are shorter
than the observation time (\ie $M \,\lsim\,100 \,M_\odot$), then the expected
total number of events is
$$N_\ev = {2\over \pi} N_* \, \tau \, \omega \, T_\eff,\eqn\nev$$
where $T_\eff$ is the effective time of the observations, which
depends weakly on the mass scale of the \ms.

The LMC will be observed for about 8 months per year.
If the \m\ events last more than
a few months (\ie \ $M \, \gsim \, M_\odot$), and
therefore bridge the four month hiatus,
then $T_\eff\sim 4\,$yrs, the length of the \m\ experiment.  For smaller
\ms, $T_\eff = 4\times 8\,$months = 2.7 yrs.  If the events last less
than a few days (\ie  $M \, \lsim \, 10^{-3}M_\odot$), then the light curves
will be too poorly sampled to be recognized as lensing events.
In this case, only
``spikes'' will be detected.  Spikes are high-magnification single-observation
events from a light curve that is too short and poorly sampled to be
resolved temporally.  Spikes can be distinguished from observational errors,
however,
because the stars are observed simultaneously in two bands.  A genuine
spike will appear equally magnified in both bands while an error (such as
a cosmic ray event) will not.  In order to be distinguished from
background, however,
the spike will need to be magnified by a factor of at least a few;
we will adopt an average minimum detectable magnification of
$A_\min = 2$.
The number of such spike events is
$$\eqalign{
N_\spike & = 2N_*\tau \bigl[ ( 1 - A_\min^{-2} )^{-1/2} - 1 \bigr] N_\obs \cr
         & \simeq N_*\tau A_\min^{-2} N_\obs \sim 135 N_*\tau  \cr
}\eqn\nspike$$
where $N_\obs$ is the
number of observations made for each of the $N_*$ program stars.
In making this evaluation,
we assumed $55\%$ good weather over 8 months of observation per year,
for a total of 4 years (K.\ Griest 1992, private communication).
We
caution that this estimate of the number of spike
events is rather uncertain due to difficulties in estimating $N_\obs$;
due to field crowding, the number of
photometric observations is not, in general, a linear function of time.
Furthermore, the value of $A_\min$ varies from star to star
with photometric errors.

While the statistical properties of the observed
spikes would be a fairly convincing signature of lensing, one would want
additional confirmation.  This could be obtained by
observing a subset of the fields, say 10\%, more frequently, say 10 times
per night.
In this way, the
individual light curves could be observed in detail.  This
altered observing schedule, however, would not substantially change
the estimate of the number of recognizable events given by equation \nspike,
but only the confidence with which spike events could be associated with
microlensing events.

\section{Null Result}

Suppose that no candidate events were observed during the four years
of observations, or that candidates were observed, but were shown to be
non-lensing events.  What classes of \m\ dark matter models could then
be ruled out?
The answer depends on the flattening of the halo and
whether or not it is truncated.

Consider first untruncated spherical halos.
 From Table 1, we find that for models with a light disk and spheroid,
the optical depth toward the LMC is
\tlmc\ $\gsim \, 3.2\times 10^{-7}$, and is somewhat greater toward
the SMC.
Since $N_* \, \gsim \, 10^7$ in both Magellanic Clouds combined (K.\ Griest
1993, private communication), we obtain $N_* \, \tau\gsim \, 3.5$.  Thus,
even for very massive \ms,
for which $\omega T_{\rm eff} \sim 1$,
at least 3.5 events are expected.  If none
were observed, this would rule out this class of \m\ models with good
confidence, that is, at the $1-e^{-3.5}\sim 97\%$ level.  From Table 2,
we find that for models with a heavy disk and spheroid, the expected
number of events may fall as low as 1.8.  Very massive \ms\ could then
be ruled out with only modest confidence, $\sim 83\%$.  However, a more
detailed analysis, taking into account the increased sampling of long
duration events, shows that moderately massive $\lsim 100 M_\odot$
\ms\ could be ruled out with good confidence.

If the halo is assumed to be spherical and
highly truncated ($R_T=2 R_0$), then
we see from Table 3 that the optical depth toward the LMC is
reduced by a factor of about $0.6$.  If the halo is assumed to
be highly flattened (E6) as well as highly truncated, then the optical
depth is reduced by another factor of about $0.75$.
In this
extreme case, the expected number of events would be about half the
number in a standard spherical halo.  If, in addition,
the disk and spheroid were heavy,
then the expected number of events would fall below unity, in which case
no definite statement could be made about heavy \ms.
 From equations \machotime\ and \nev, we see that then
only \ms\ of mass $M \, \lsim \, {\cal O}(10)\,M_\odot$
would be ruled out at a 97\% confidence level.

In short, a null result for the \m\ Collaboration experiment
would rule out a standard (spherical and untruncated) \m\ halo for
masses $M \, \lsim \, 10^3$ - $10^6 M_\odot$, depending on the mass of
disk as measured by the \m\ experiment itself.  However, a null result
would rule out a highly truncated and highly flattened halo only
for $M \, \lsim \, {\cal O}(10-100) \,M_\odot$.
The higher limit of $100\,M_\odot$ assumes that very long duration events
will be so well sampled that events of lower amplification
will contain enough information to discern
the signature of microlensing.

\section{Measurement of Halo Flattening}

If \ms\ are detected then, as discussed in \S\S\ 2, 4, and 5, the
ratio of the optical depths toward the SMC and LMC will be a robust
indicator of halo flatness.  The precision with which this ratio
can be measured depends on the Poisson fluctuations in the number
of events detected, and the lensing ``backgrounds'' toward the
two Magellanic Clouds.
We address each of these in turn.

For the purposes of making our estimates of the statistical fluctuations,
we assume that about 20\% of the Magellanic program stars
will be in the SMC, so that $N_{*,\smc}\sim 2\times 10^6$.  Most of
the statistical fluctuation in the measurement of \tsmctlmc\
will then from \tsmc.
The optical depth toward the SMC varies somewhat from
model to model, but for these purposes we
adopt \tsmc\ $\sim 4\times 10^{-7}$.  Thus,
$N_{*,\smc} \, $ \tsmc\ $\sim 0.8$.  From equations \nev\ and \nspike, we
therefore estimate the fractional precision of the ratio measurement as
$$\eqalign{\sigma\biggl(\ln{\tau_\smc\over\tau_\lmc}\biggr)\sim &
\Bigl(\min\{N_{\ev,\smc},N_{\spike,\smc}\}\Bigr)^{-1/2}\cr
\sim & 10\% \times \max\{1,(M/ 0.004 M_\odot)^{1/4}\}}.\eqn\sigest$$
Since \tsmctlmc\ varies by a factor $\sim 1.5$ between
E0 and E6 halos, it should be possible to distinguish between
these two cases, provided that the \ms\ are substellar.

The principal lensing ``background'' comes from events due to \ms\
in the Magellanic Clouds themselves.  The LMC has a modest
halo characterized by a rotation speed of about $80\,\kms$
(Schommer \etal 1992);
this halo may or may not be composed of \ms.  The optical depth of the
SMC halo (if it exists) is an order of magnitude smaller than that
of the LMC and can be ignored
here.  (See Gould 1993 for
an extensive discussion of lensing by \ms\ in the LMC and SMC.)\ \
The optical depth due to \ms\ in a spherical LMC halo is
$\tau_{\lmc,\rm int}\sim 10^{-7}$.  For a flattened LMC halo, the optical
depth is reduced by approximately the flatness ratio, $\cos\psi_\lmc$.
Hence, the fraction of all lensing events toward the LMC that are due to LMC
\ms\ is not
necessarily
negligible, and the {\it a priori} uncertainty in this fraction
may be as much as 10 or 15\%.

It is possible to reduce this uncertainty
in two ways.
First, the difference in the optical depths toward the
far and near sides of the LMC would allow one to measure $\tau_{\lmc,\rm int}
\cos\psi_\lmc$ (Gould 1993),
which, unfortunately, is not quite the quantity of most interest,
$\tau_{\lmc,\rm int}$.
Moreover, the planned four years of observations
by the \m\ Collaboration will not be sufficient to make a precise
measurement of this quantity.
Nevertheless, even the crude constraint obtained from
this measurement would be useful.  Second, by making several
observations per day of all lensing events
identified to be in progress, it would be
possible to measure directly and with good precision the fraction of
events due to LMC Machos (Gould, in preparation).
These additional observations would allow one to pick out those
lensing events in which the \m\ passed directly over (or very near)
the face of a lensed star.  For these events, one could measure the
angular speed of the \m\ from the deviation of the light curve
from a standard ``point-source'' light curve.  Since the angular
speeds of LMC \ms\ are about 15 times slower than those of Galactic
\ms, LMC \ms\ could be recognized unambiguously.  In this way,
the optical depth of LMC \ms\ could be measured accurately.
Unfortunately, the organization of these additional
observations would be a major undertaking.

In brief, there is a systematic uncertainty of $\lsim 15\%$
in the measurement of the ratio \tsmctlmc\ due to the
``background'' of lensing by LMC \ms.  This uncertainty can be reduced
somewhat by analyzing the spatial distribution of lensing events and
can be greatly reduced by undertaking significant additional observations.
The statistical uncertainty of the measurement of \tsmctlmc\
is $10\%$ provided that the \ms\ are substellar.

\chapter{Conclusions}

The primary conclusions of this work can be summarized in four main
points:

$\bullet$ If the dark halo of the Galaxy is composed of \ms, the ratio
of optical depths toward the SMC as
compared to the LMC, \tsmctlmc, is a robust measure of the ellipticity
of the halo, independent of any current limitations on our knowledge of
the the mass of the Galactic disk and spheroid, the slope of the Galactic
rotation curve, and the truncation radius of the dark halo.

$\bullet$ If \ms\ are substellar and the SMC is sufficiently sampled,
the ongoing \m\ Collaboration experiment
will be sensitive enough to use \tsmctlmc\
to measure the flattening of the Galactic dark halo.

$\bullet$ The optical depth to the Galactic bulge is {\it not\/} an effective
probe of the shape of the dark halo because the expected rate of
microlensing is sensitive to Galactic parameters
(\eg the mass of the disk and spheroid), and because contamination due
to disk and bulge stars make any events due to \ms\ difficult
to disentangle.

$\bullet$ For extreme assumptions about the shape of the Galactic dark
halo (E6), the halo truncation radius (twice the solar circle), and the
mass of the Galactic disk ($\Sigma_0=75\,M_\odot\,\rm pc^{-2}$) and
spheroid ($2.4 \times 10^{10} \msolar$), the optical depth toward the LMC
is sufficiently reduced that the upper limit on the \m\ mass range
to which ongoing experiments are sensitive would be lowered to
${\cal O}(10)\,M_\odot$.
This contrasts sharply with the upper limit of
${\cal O}(10^6)\,M_\odot$ for more standard Galactic parameters.

\vskip 4in

{\bf Acknowledgements:}
We would like to thank Charles Alcock for a careful reading of the
manuscript resulting in many useful comments,
and Hans-Walter Rix for valuable discussion.
Work by AG was supported by the National Science Foundation
(PHY~92-45317).  PDS was supported by the National Science
Foundation (AST~92-15485) and the J. Seward Johnson Charitable Trust.

\vfil\eject

\centerline{\bigfont Appendix:~~Inversion of the R$^{^{\scriptstyle 1/4}}$
Law to Obtain}
\centerline{\bigfont the Volume Density and Force: Spherical Case}

{}~

A derivation of the spherical volume density distribution
consistent with a de~Vaucouleurs ($R^{1/4}$ law) surface mass density
was given in integral form by Poveda
\etal (1960).
Young (1976) used Poveda's expression to tabulate numerically the
enclosed mass, mean density, force, potential, and escape velocity as a
function of radius in the galaxy.
A closed-form, analytic approximation to the integral form has been given
by Mellier \& Mathez (1987); two parameters of their analytic form must
be varied with galactocentric radius in order to reproduce the $R^{1/4}$
law over a large radial range.
Here, we use Young's normalized form
to derive an expression for the
mass interior to r, which is needed to calculate the gravitational
force due to the spheroid at any point.  Since the original reference can be
difficult to find, we first rederive the form of the density, $\rho(r)$,
that gives an $R^{1/4}$ spheroid in projection, and show that it reduces
to Young's normalized form.

The projected surface density (assuming $M/L\/$ is constant) of a
$R^{1/4}$ spheroid is given by
$$\Sigma(R) = \Sigma(0)~\exp{ \biggr[ - b \ \biggr({R \over
R_e}\biggl)^{1/4} \biggl] }\ ,\eqn\dVsmd$$
where $R$ is the projected galactocentric radius on the sky, $R_e$ is the
effective radius that encloses half the projected mass (or light), and
$b=7.66925$.  We require that the spherical density distribution $\rho(r)$
give this form for the surface mass density in projection:
$$\Sigma(R) = \int_{- \infty}^{\infty} dz~\rho(r) = 2 \int_{R}^{\infty}
 dr~{\rho(r) \over \sqrt{ 1 - (R/r)^2}}\ . \eqn\dVsmdrint$$

If we change to the normalized variables $y \equiv (r/R_e)^2$ and
$x \equiv (R/R_e)^2$, then
$$\tilde\Sigma(x) = \Sigma(0) \ \exp{(-b x^{1/8})} = R_e \int_x^{\infty} dy
{\tilde\rho(y) \over \sqrt{y - x}} \ . \eqn\dVsmdn$$
We can now use an Abel Transform (see \eg Binney \& Tremaine 1987) to
derive an integral expression for $\tilde\rho(y)$.

For any functions $f(x)$ and $g(y)$ satisfying
$$f(x) = \int_x^{\infty} dy \ {g(y) \over (y - x)^{1-\alpha}}\ ,
{}~~~~~~~~~~~~~~{\rm where} ~~~~0 < \alpha < 1,\eqn\fx$$
$g(y)$ is given by
$$\eqalign{
g(y) &= {- {\sin \pi \alpha} \over \pi} {d \over dx}
\int_{y}^{\infty} dx \ {f(x) \over (x - y)^{1 - \alpha}}  \cr
          &= {- {\sin \pi \alpha} \over \pi} \biggl[
\int_{y}^{\infty} \biggr( {df(x) \over dx} \biggl) \
{dx \over (x - y)^{1 - \alpha}}
- \lim_{x \to \infty} {f(x) \over (x - y)^{1 - \alpha}} \biggr]   \cr
}\eqn\gy$$

Thus, applying the Abel Transform to equation \dVsmdn\ yields
$$R_e \ \tilde\rho(y) = {{b \ \Sigma(0)} \over \pi} \ \int_{y}^{\infty}
 dx~{ {x^{-7/8} \ \exp{[-b x^{1/8}]}} \over {8 \ \sqrt{x - y}}}\ .\eqn\rerho$$
To retrieve the form given by Poveda \etal (as recast by Young), we must first
change variables
$$ s^2 \equiv (r/R_e)^2 \equiv j^8 = y~,~~~~~~x/y = t^8 $$
$${\rm so~~that}~~~~~\rho(r) = {{b \ \Sigma(0)} \over {\pi \ R_e \ j^3}}
\int_{1}^{\infty}
 dt~{ e^{-b j t} \over {\sqrt{t^8 - 1}}} \eqn\rhointermed$$
and then rewrite the normalization in terms of the total mass, $M_T$.
Integrating the surface mass density of an
$R^{1/4}$ profile given by eq.\ \dVsmd, we find
$$\Sigma(0) = { {M_{T} \ b^8} \over {8! \ \pi \ R_e^2} } \ ,$$
which implies
$$\rho(r) = { {M_{T} \ b^9} \over {8! \ \pi^2 \ R_e^3 \ j^3} }
\int_{1}^{\infty} dt~{ e^{-b j t} \over {\sqrt{t^8 - 1}}} \eqn\rhor$$

This form should be compared to that of Young, whose dimensionless
density is given as
$$\tilde\rho(s) = { 1 \over {2 \ j^3} }
\int_{1}^{\infty} dt~{ e^{-b j t} \over {\sqrt{t^8 - 1}}} \eqn\rhos$$
and whose the dimensionless total mass, $\tilde M$, is defined as
$\tilde M \equiv \int_{0}^{\infty} ds \ \tilde\rho(s) \ 4\pi \ s^2 \ $

 From these definitions of $s$ and $\tilde M$, then,
$$\rho(r) = \biggr( {M_T \over {\tilde M \ R_e^3}} \biggl) \tilde \rho(s)
\eqn\rhorel$$
which requires, by comparison of equations \rhor\ with \rhos, that
$$\tilde M = { {8! \ \pi^2} \over {2 \ b^9} }\ , \eqn\mtildeval$$
This analytic result agrees with the numerical result of Young who found that
$\tilde M = 2.1676 \times 10^{-3}$.

Our primary interest is to compute the gravitational force due
to a spherical de~Vaucouleurs spheroid at any distance $r_0$, so that we
can compare the expected circular rotation speed for a given mass
model with an assumed Galactic rotation curve.  To this end, we require
an expression for
the mass from $\rho(r)$ enclosed by a sphere of radius $r=r_0$.
 From equation \rhorel\ we have
$$\eqalign{
M(< r_0) &= {{4\pi \ M_T } \over \tilde M} \int_0^{r_0/R_e} ds \
\tilde \rho(s) \  s^2 \cr
         &= {{8\pi M_T} \over \tilde M} \int_1^{\infty} {dt \over \sqrt{t^8-1}}
\ \int_0^{(r_0/R_e)^{1/4}} dj \ j^8 \ e^{-b t j}  \cr
}\eqn\menc$$
By repeated integrations by parts, the integral over $j$ can be done
analytically to obtain a polynomial in $(bt)^{-1}$ that multiplies
an exponential factor,
$$M(< r_0) = {{16 M_T} \over \pi}
\int_1^{\infty} {dt \over {t^9 \, \sqrt{t^8-1}} }
\biggl\{ 1 - e^{ -bt (r_0/R_e)^{1/4} } \sum_{n=0}^8 {1 \over n!}
\bigl[bt \, (r_0/R_e)^{1/4} \bigr] ^n \biggr\} ~ .
\eqn\mencseries$$

One is then left with the
integral over dt, which can be done numerically.  Since
$(t^8-1) = (t-1)(t+1)(t^2+1)(t^4+1)$, the integrand does have an
integrable square root singularity at the lower limit, which requires
some care.  The radial acceleration at $r$ is then just given by
$a = G M(<r)/r^2$.

\endpage

\Ref\MC{Alcock, C.,\ \etal\ 1992, in {\it Robotic Telescopes In The 1990s},
ed.\ A. Filipenko, (San Francisco: Astronomical Society of the Pacific)}
\Ref\Ash{Ashman, K., \ 1992, PASP, 104, 1109}
\Ref\Au{Aubourg, E.,\ \etal\ 1991,  Pre-publication 346,
Presented at 2nd DEAC meeting, Observatoire de Meudon, France 5-8 March}
\Ref\Bah{Bahcall, J.~N.\ 1984, ApJ, 276, 169}
\Ref\BS{Bahcall, J.~N., \&  Soneira, R.~M.\ 1980, ApJ~Supp, 44, 73}
\Ref\BFG{Bahcall, J.~N., Flynn, C., \& Gould, A.\ 1992, ApJ, 388, 345}
\Ref\KGB{Begeman, K.~G.\ 1987, Ph.D. Thesis, University of Groningen}
\Ref\BT{Binney, J.\ \& Tremaine, S.\ 1987, {\it Galactic Dynamics} (Princeton:
Princeton University Press)}
\Ref\BK{Blandford, R.~D.\ \& Kochanek, C.~S.\ 1987, {\it Dark Matter in the
Universe: Proceedings of the Fourth Jerusalem Winter School for Theoretical
Physics} eds.\ J. Bahcall, T. Piran, and S. Weinberg, (Singapore: World
Scientific) p.\ 139}
\Ref\DC{Dubinski, J. \& Carlberg, R.~G.\ 1991, ApJ, 378, 496}
\Ref\FBS{Fich, M., Blitz, L. \& Stark, A.~A.\ 1989, ApJ, 342, 272}
\Ref\FT{Fich, M. \& Tremaine, S.\ 1991, ARAA, 29, 409}
\Ref\Gr{Griest, K.,\ \etal\ 1991, ApJ, 372, L79}
\Ref\Gouldzero{Gould, A.\ 1990, MNRAS, 244, 25}
\Ref\Gouldtwo{Gould, A.\ 1992, ApJ, 392, 442}
\Ref\Gouldthree{Gould, A.\ 1993, ApJ, 404, 451}
\Ref\Cats{Katz, N.\ 1991, ApJ, 368, 325}
\Ref\CatsandGunns{Katz, N. \& Gunn, J.~E.\ 1991, ApJ, 377, 365}
\Ref\Ken{Kent, S.~M.\ 1987, AJ, 93, 816}
\Ref\kga{Kuijken, K.\ \& Gilmore 1989, MNRAS, 239, 605}
\Ref\TL{Lauer, T.~R.\ 1985, ApJ, 292, 104}
\Ref\Merry{Merrifield, M.~R.\ 1992, AJ, 103, 1552}
\Ref\Pac{Paczy\'nski, B.\ 1986, ApJ, 304, 1}
\Ref\Pact{Paczy\'nski, B.\ 1991, ApJ, 371, L63}
\Ref\RP{Peletier, R.\ 1989, Ph.D. Thesis, University of Groningen}
\Ref\Poveda{Poveda, A., Iturriaga, R., and Orozco, I.\ 1960, Bol. Obs.
Tonantzintla, 20, 3}
\Ref\SS{Sackett, P. D. \& Sparke, L.S., \ 1990, ApJ, 361, 408}
\Ref\PS{Schechter, P.\ 1993, in ``Back to the Galaxy,'' proceedings of
STScI workshop, ed.~F.~Verter}
\Ref\SOSH{Schommer, R.\ A., Olszewski, E.\ W., Suntzeff, N.\ B., \&
Harris, H.\ C.\ 1992, AJ, 103, 447}
\Ref\ogle{Udalski, A., Szyma\'nski, J., Kaluzny, J., Kubiak, M., \&
Mateo, M.\ 1992, Acta Astronomica, 42, 253}
\Ref\Young{Young, P.~J.\ 1976, AJ, 81, 807}
\Ref\Zari{Zaritsky, D., Smith, R., Frenk, C. \& White, S.~D.~M.\ 1992, ApJ,
405, 464}
\refout

\vfil\eject

\centerline{FIGURE CAPTIONS}
\bigskip
\noindent {\bf Fig.~1~~}
The optical depth to microlensing as
a function of halo core radius $a$ in kpc
for a spherical (E0, left panels) and flattened (E6, right panels)
Galactic dark halo, along lines of sight
to the SMC (top panels) and LMC (bottom panels).  The halo parameter
$v_\infty$ (see text) is assumed to be 220 \kms\ for the purposes of
this figure; all optical depths scale strictly as $v_\infty^2$.
Solid lines trace the optical depths through an untruncated
Galactic halo (that is, a halo that extends to at least the Magellanic
Clouds);
dashed lines are for a halo truncated in the Galactic plane at four
times the solar circle; dotted lines indicate a halo truncated at
twice the solar circle.
\bigskip
\noindent {\bf Fig.~2~~}
The ratio of optical depths along SMC and LMC
lines of sight, \tsmctlmc, as a function of halo core radius
for a spherical (E0) and flattened (E6)
Galactic dark halo for different halo truncations.  Note the expanded
vertical scale; the ratios vary only weakly with core radius, but are
strong functions of halo flattening.
\bigskip
\noindent {\bf Fig.~3~~}
The ratio of optical depths along SMC and LMC
lines of sight, \tsmctlmc, as a function of assumed Galactic
rotation curve for four different mass models for the Galactic disk
and
spheroid, and three different halo truncation radii.
(See the text for a detailed description of the models.)
The rotation curve
models are numbered as in Tables 1, 2 and 3, namely: $1 = {\rm
fall-fall}$,
$2 = {\rm fall-flat}$, $3 = {\rm fall-rise}$, $4 = {\rm flat-fall}$,
$5 = {\rm flat-flat}$,
$6 = {\rm flat-rise}$, $7 = {\rm rise-fall}$, $8 = {\rm rise-flat}$,
$9 = {\rm rise-rise}$.
Note that for a given halo flattening and truncation radius, all four
mass models and all nine rotation curve models give nearly identical
\tsmctlmc\ ratios.  Only the heaviest disk and spheroid model deviates
somewhat for rotation curve models 6, 8, and 9.  The ratio \tsmctlmc\
is more sensitive to halo flattening than any other Galactic
parameter,
including halo truncation.
\end